\documentclass[groupedaddress,floatfix]{revtex4}
\newsavebox{\HttpBox}
\newcommand{\mul}[1]{\multicolumn{1}{c}{#1}}
\newcommand{\mult}[1]{\multicolumn{2}{c}{#1}}
\begin{lrbox}{\HttpBox}
\footnotesize\verb+http://www.mpip-mainz.mpg.de/~yasp/ccteam/index-gruppe.html+
\end{lrbox}
\newsavebox{\EBox}
\begin{lrbox}{\EBox}
\footnotesize\verb+rfaller@ucdavis.edu+
\end{lrbox}
\usepackage{amsmath}
\usepackage[dvips]{epsfig}
\usepackage{dcolumn}
\newcommand{\mc}[1]{\multicolumn{1}{c}{#1}}
\newcommand{\opC}{\operatorname{C}}
\newcommand{\opH}{\operatorname{H}}
\begin{document}
\title{Properties of Poly (isoprene) - Model Building in the Melt and in
  Solution}
\author{Roland Faller\footnote{Electronic Mail: \usebox{\EBox}}}
\affiliation{ Department of Chemical Engineering \& Materials Science,
University of California-Davis, Davis, CA 95616, USA} \author{Dirk
  Reith\footnote{Present address: DaimlerChrysler
    AG, HPC G202, Fronaeckerstr.\ 40, D-71059 Sindelfingen, Germany }}
\affiliation{Max-Planck-Institut f\"ur Polymerforschung, 55128 Mainz, Germany}
\begin{abstract}
\noindent Properties of 1,4-\textit{trans} poly (isoprene) at ambient
conditions are determined by simulations on two length scales based on two
different models: a full-atomistic and a mesoscopic one. The models are linked
via a mapping scheme such that one mesoscopic bead represents one chemical
repeat unit. Melts as well as solutions of several chain lengths were
investigated and mapped individually to the meso-scale. The resulting models
are compared to each other. The meso-scale models could be simulated over a
large variety of chain lengths and time-scales relevant for experimental
comparison. Concerning static properties, we determined the persistence length
of our systems and the scaling behavior of the radius of gyration. The latter
was compared to experiments and the agreement is satisfactory. Furthermore, we
find deviations from Rouse dynamics at all chain lengths at ambient conditions.
\end{abstract}
\maketitle
\section{Introduction}
Poly (isoprene), better known as rubber, is one of the most commonly
used polymers.  Its natural form, kautschuk, is highly abundant in
nature and for technological purposes, synthetic variations are easily
industrially polymerised from isoprene~\cite{stevens99}. As an
elastomer, its rich field of technical applications comprises tires,
textiles, or cable coatings~\cite{cowie97}.  Still, some of its static
as well as dynamic properties are still not understood on a
microscopic basis.

The time- and length-scales important for the investigation of
polymeric materials generally cover a wide range - from Angstroms and
picoseconds to millimeters and eventually hours. There is no hope to
capture all important quantities of a system with one single
model. Thus, computational methods have to be developed to use the
information of more detailed models as an input for coarser models
allowing to investigate larger systems.  The most prominent example is
the step from quantum chemistry (based on Schroedinger's equation) to
semi-empirical atomistic models (based on Newton's equation of
motion). For many modern applications, however, this step is not
sufficient. So, the mapping of atomistic systems to meso-scale systems
(in which monomers, thought of as "super-atoms", are the smallest
units of the simulation) became the focus of several studies.

Coarse-graining (CG) approaches have reached considerable attention
over the last decades~\cite{baschnagel00a}. The basic idea of all CG
schemes is to separate the system into fast and slow variables where
the fast degrees of freedom are not of primary interest for the
question under study. Thus, a model can be compiled containing the
slow degrees of freedom only. Recently methods have become available
to perform coarse graining in a polymer-specific and systematic
way~\cite{tschoep98a,reith00s,reith01se,akkermanns01}. We refer to a
systematic procedure if the identity of the polymer is not lost during
the mapping procedure. Moreover, the purpose of such techniques is to
develop meso-scale models for the simulation of large systems of
specific polymers. To ease the mapping procedure, two automatic
optimization schemes have been developed by our coworkers and us: firstly
a simplex algorithm technique~\cite{meyer00,reith02s}, and secondly the
iterative Boltzmann inversion method which directly targets the problem of
structure differences with a physically motivated
approach~\cite{puetzreith01e}.  Similar ideas have been applied to
reproduce experimental data in atomistic simulations~\cite{faller99c}.

The goal we pursue here is that two models on different length scales should
produce the same distribution functions on the larger of the two length scales.
Then, we can regard them representing the same polymer. Due to its rich
applications, we take poly (isoprene) as an example. Additionally we point out
some differences to recent findings in studies of semi-generic bead-spring
models incorporating stiffness as the only local
characteristics~\cite{faller01b,faller02a}.
\section{Investigations with atomistic detail}
As a basis for our mapping procedure we need atomistic simulations with
good accuracy. We performed simulations with appropriately designed
force-fields reproducing the atomistic structure and the thermodynamics very
reliably.
\subsection{Melt Investigations}
\begin{figure}
  \begin{center}
    \includegraphics[width=8.25cm]{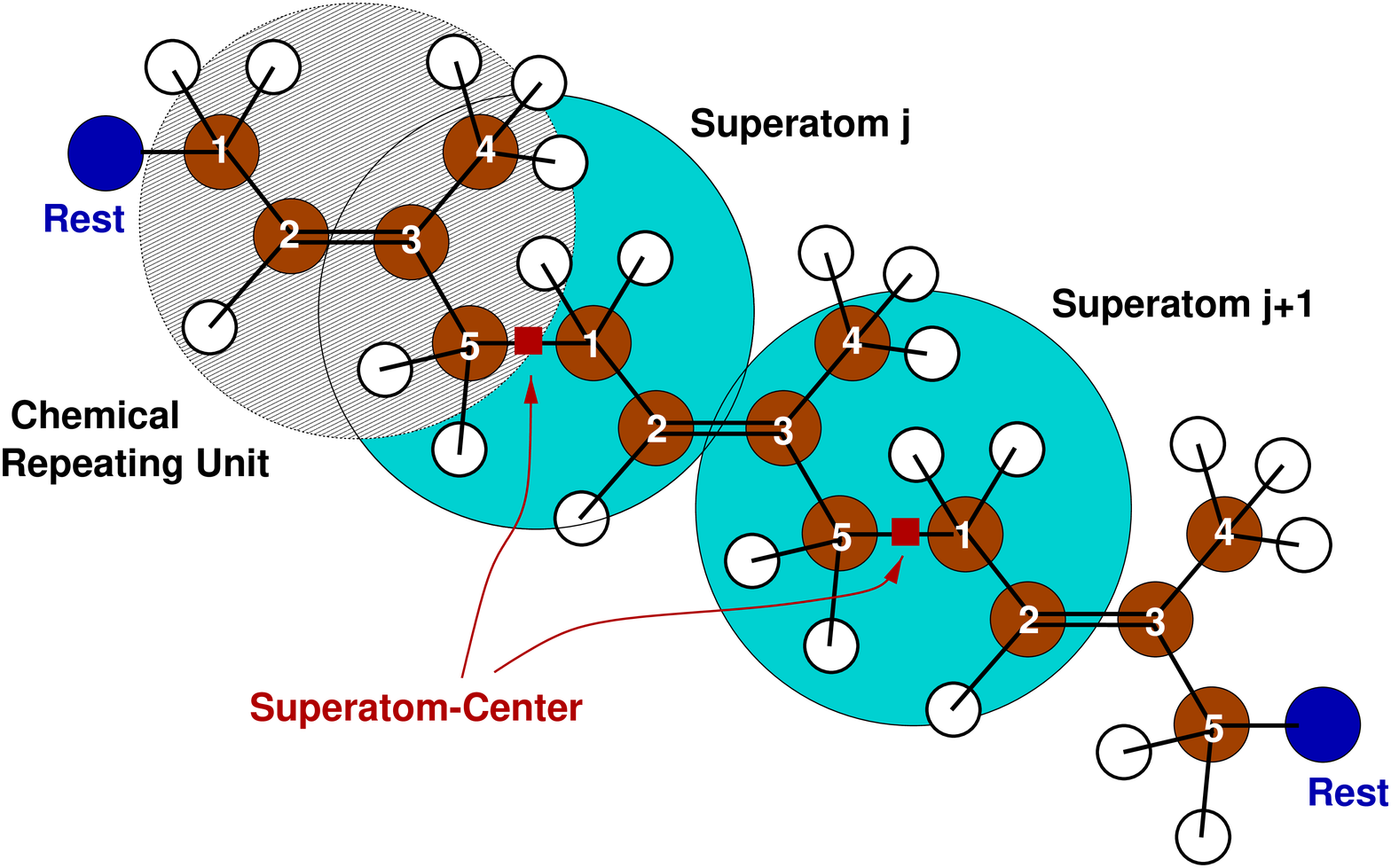}
  \end{center}
  \caption {Illustration of the mapping of
    \textit{trans}-1,4-poly (isoprene) from the atomistic to the mesoscopic
    level. Each chemical repeat unit is represented by one super-atom. As
    center of these super-atoms, we choose the middle of the atomistic bond
    between two successive chemical repeat units. This is useful because the
    resulting mapped chains generate well distinguishable peaks for the
    various intramolecular distributions.}
  \label{polyisoprene}
\end{figure}

The atomistic melt simulations have been described in detail in
reference~\cite{faller01a}. Here, we briefly summarize the main
characteristics. The simulation box contains 100 oligomers of average length 10
monomers. All chains represent {\it trans}-poly (isoprene)
(cf.\ Figure~\ref{polyisoprene}). We use a self-developed all-atom force-field
resulting in 132 interaction sites for a 10-mer~\cite{faller01a}; every atom is
represented by one interaction site
(cf. Tables~\ref{tab:ffangbnd}-\ref{tab:torspot}). The simulations lasted for
1 to 2~ns at ambient conditions ($T=300$~K, $p=101.3$~kPa), cf.
Table~\ref{tab:simlen}. The model is capable of describing local reorientation
functions in comparison to NMR measurements and reproduces reasonably the melt
structure factor of poly (isoprene)~\cite{faller00a,faller01a}.
Atoms connected by any bonding potential did not interact by the Lennard-Jones
potential. Additionally, the following non-bonded interactions were excluded:
all within one monomer, and all C$-$C, C$-$H and H$-$H interactions between
the second half of the carbons of one monomer (atoms C$_3$, C$_4$, and C$_5$)
and the first half of its following neighbor (C$_1$ and C$_2$) including the
hydrogens connected to them.

Constant temperature and pressure are ensured using Berendsen's
method~\cite{berendsen84} with time constants 0.2~ps for temperature and 8~ps
for pressure, respectively. The pressure coupling using a compressibility of
$2\times10^{-7}~\text{kPa}^{-1}$ was employed for the three Cartesian
directions independently. All simulations were performed using the {\sc YASP}
molecular simulation package~\cite{yasp} with a time-step of 1~fs and a cutoff
for the non-bonded interactions at 0.9~nm. Configurations were saved every
picosecond. No charges were used through the simulations.
\subsection{Atomistic Simulations of Poly (isoprene) in Solution of
Cyclohexane}
For the atomistic simulations of {\it trans}-PI in cyclohexane we used the same
force-field as in the melt simulations except that only interactions up to
$1-4$ interactions (inclusive) are excluded. The cyclohexane force-field was
developed using the automatic simplex method~\cite{faller99c} and tested in the
dynamics of mixtures of cyclohexene and cyclohexane~\cite{schmitz99a}. For
this study we augmented the cyclohexane force-field with torsion potentials
along the ring in order to avoid indavertent flipping between the chair and
boat configuration~\cite{faller02d}.
The densities of the solutions increase slightly with concentration of polymer
(Tab.~\ref{tab:systems}). They are very close to the density of pure
cyclohexane without torsions (767.7~kg/m$^3$)~\cite{faller99c,schmitz99a},
which in turn was optimized against experimental data. We did not reoptimize
the Lennard-Jones parameters.  Therefore the pure cyclohexane system has a
slightly lower density as the molecules are less flexible and less spherical
due to the torsions. In order to decide on the quality of the solvent the
radius of gyration
\begin{equation}
  R_G = \sqrt{\sum_i(m_ir_i^2) / \sum_i m_i },\;
        \vec{r}_i = \vec{R}_i - \langle \vec{R}_i\rangle
\end{equation}
and the end-to-end length are important characteristics. The mean square
end-to-end distance $\sqrt{\langle R_{\text{e-e}}^2\rangle}$ is measured
between the terminal carbons of the chain. Additionally the hydrodynamic
radius $R_H$ was calculated as the first inverse moment of the distance
vectors~\cite{doi86}:
\begin{equation}\label{eq:rhydro}
  \frac{1}{R_H} \;  = \; \frac{1}{N^2}%
  \left\langle \sum_{i\neq j}\frac{1}{r_{ij}}\right\rangle \; .
\end{equation}
Here, $N$ represents the number of monomers of the chain and $r_{ij}$ the
distance between two arbitrarily chosen super-atoms $i$ and $j$.

We do not find a systematic variation of the radius of gyration with
concentration. In the system with two chains in a solvent of 500~molecules the
two chains behave differently whereas in the higher concentrated systems the
radii of gyration are very close although they were also started with the same
initial configurations as the other two chain system. The auto-correlation
functions of the gyration radii, however, decays on the scale of less than a
hundred pico-seconds. Thus, the systems are equilibrated even though the error
bars are quite large. In order to decrease them substantially much larger
simulations would be necessary. This was not the purpose of the atomistic
simulations as the long time-scales are left to the coarse-grained model to be
developed here.

\begin{figure}
  \begin{center} \includegraphics[width=8.25cm]{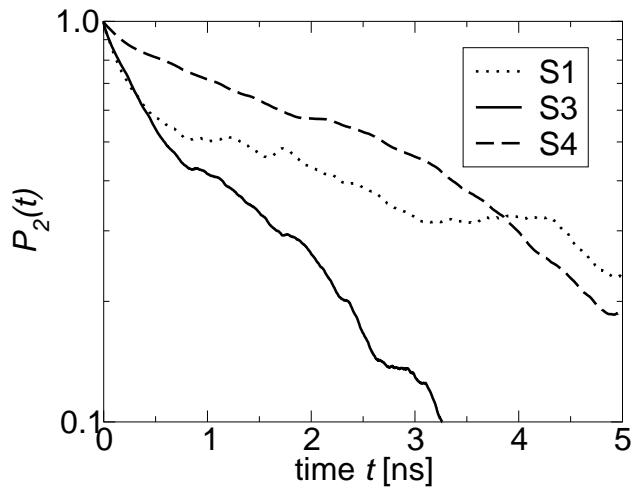}
  \end{center} \caption {Reorientation correlation function of whole
  chains in solution in semilogarithmic representation. We show the
  second Legendre polynomial $P_2$ of the reorientation angle. The
  shortcuts S1, S3, and S4 are defined in Table~\ref{tab:systems}.}
  \label{fig:e-e}
\end{figure}

In Figure~\ref{fig:e-e} we show the decay of the reorientation correlation
function of the end-to-end vector of the whole chains in solution. We see that
this function overall follows an exponential tendency and the correlation time
is in the order of $2-4$~ns. This supports once again the assumption
of equilibration. Additionally the overall reorientation becomes slower with
increasing concentration. The dynamics of the solvent is found to slow down and
become more anisotropic with increasing concentration~\cite{faller02d}.
\section{The Mapping Procedure}
The mapping procedure is the same for both the melt and the solution system.
It has already been described in full detail in reference~\cite{puetzreith01e},
so we limit ourselves to a short summary. Distribution functions from
atomistic simulations are the basis of the mapping, since the mesoscale model
is optimized against them. In this study, M1 (for the melt) and the combination
of S1 and S3 (for the solution) were chosen as parent atomistic simulations.
The mapping itself is illustrated in Figure~\ref{polyisoprene}. Each chemical
repeat unit is replaced by one super-atom (interaction center), located in the
middle of the atomistic bond between successive chemical repeat units.
Consequently an atomistic $N$-mer will be coarse grained to a $N-1$-mer. This
is meaningful because the such mapped super-atoms are much more spherical
compared to the case in which one would have chosen the center of mass of the
real monomer. The advantage is that the mapped chains generate well
distinguishable peaks for the various intramolecular distributions. Especially
when applying the model to longer chains, the error introduced due to the
missing rests are negligible.

In this contribution, we use exclusively the pressure-corrected CG force field
which has been optimized by the Inverted Boltzmann method. It is fully
described in Reference~\cite{puetzreith01e}. The idea, however, shall shortly
be illustrated here. The Inverted Boltzmann method utilizes the differences in
the potentials of mean force between the distribution functions generated from
a guessed potential and the true distribution functions to improve the
effective potential successively. Imagine we would like to derive an effective
non-bonded potential $V_{\infty}(r)$ from a given tabulated start potential
$V_0(r)$, targeting to match the radial distribution function $g_{\infty}(r)$.
Simulating our system with $V_0(r)$ will yield a corresponding $g_0(r)$ which
is different from $g_{\infty}(r)$. The potential needs to be improved, which
is done by a correction term $- k_B T \ln \left( \frac{g_0(r)}{g(r)}\right)$.
This step can be iterated as follows by:
\begin{equation}
 V_{j+1}(r) = V_j(r) - k_B T \ln \left( \frac{g_j(r)}{g(r)}\right) \,
\end{equation}
until
\begin{equation} \label{e:meritrdf}
f_{\text{target}} = \int_{r_{\rm min}}^{r_{\rm max}} \omega(r)
            \left(g_{\infty}(r) - g_j(r)\right)^2  dr \;.
\end{equation}
falls below an initially specified threshold. (We apply $w(r)=exp(-r)$ as
weighting function in order to additionally penalize deviations at small
distances.) Pressure corrections can be introduced in polymer systems by
adding a week linear potential term to the attractive outer part of $V_j(r)$.
One can then post-optimize the structure of the system according to the above
scheme in turn with occasional linear additions until the pressure
matches the one from the atomistic system, too.
\section{Mesoscale simulations}
\subsection{Technical Simulation Details}
Both Brownian Dynamics (BD, for the melt) and Monte Carlo (MC, for the
solution) programs have been applied. All BD runs were performed in the $NVT$
ensemble, with values for density and temperature corresponding to the values
of the parent atomistic simulation. The systems consisted of an orthorhombic
box employing periodic boundary conditions. Langevin equations of motion were
integrated by the velocity Verlet algorithm with a time step $\Delta t =
0.01\tau$~\cite{allen87} at a dimensionless temperature of $T^*=1$. This
temperature was maintained by the Langevin
thermostat with friction constant $\Gamma = 0.5 \tau^{-1}$~\cite{grest86}. For
the MC simulations representing the solution case, a single chain program was
used~\cite{puetz01}. Here, $10^5$ accepted warm-up moves were carried out
before a production run of $10^6$ accepted moves was started. For analysis,
every 500th configuration was stored, both in BD and MC simulations.
\subsection{Static properties in the melt and in solution}
The results of the CG simulations are summarized in Table~\ref{tab:pip_melt}
in case of the poly (isoprene) melts and in Table~\ref{tab:pip_solution} in
case of the solutions. For both, the radius of gyration $R_G$ and the
hydrodynamic radius $R_H$ is correctly reproduced compared to the values of
the parent atomistic simulations (within standard errors).

\begin{figure}
  \begin{center}
    \includegraphics[width=8.25cm]{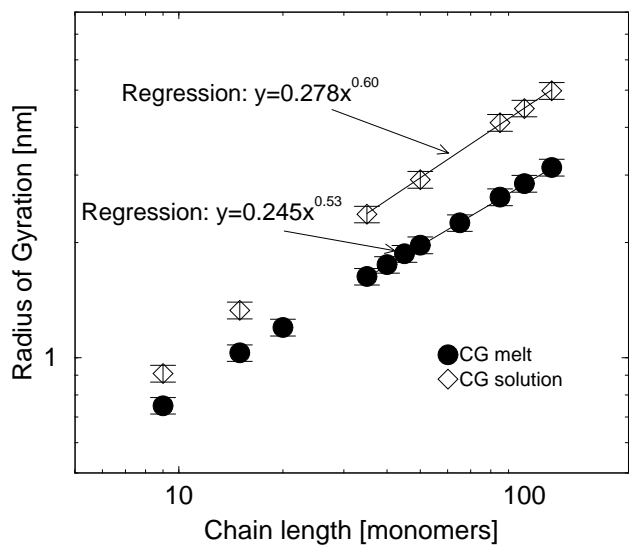}
    \caption{Scaling behavior of CG poly (isoprene) chains in melt (filled
      circles) and solution (open diamonds).}
    \label{scaling_RG}
    \end{center}
\end{figure}

Extensive Brownian Dynamics simulations were carried out to investigate the
static and the dynamic behavior of poly (isoprene) melts. The coincidence
(within statistical fluctuations) of the values for $D_{\rm center}$ and
$D_{\rm com}$, representing the diffusion coefficient of the central monomer of
the chain and their center of mass, respectively, shows that the system is
fully equilibrated.  Let us here concentrate on the static properties. In
Table~\ref{tab:pip_melt} we list the values for the radius of gyration $R_G$,
the hydrodynamic radius $R_H$, their ratio and the ratio $R_e^2/R_G^2$ which
provide information about the extension of the chains. The scaling behavior of
$R_G$ is shown in Figure~\ref{scaling_RG}. A fit for $N \ge 35$ yields a
scaling exponent of $\nu = 0.53$, which is slightly larger than the ideal value
for melts of $\nu_{\theta} = 0.5$, suggesting that the chains are not
sufficiently close the the limit of infinitely long chains. This is also
supported by the ratio $R_G/R_H$ which is well below the infinite limit of
around $1.5$. The fact that $R_e^2/R_G^2 = 6.0$ holds for all chain lengths,
however, shows that the static behavior corresponds to random walks.

Table~\ref{tab:pip_solution} holds the results of the dissolved poly (isoprene)
systems. Since we applied MC, static properties are listed exclusively. The
scaling behavior of $R_G$ is also presented in Figure~\ref{scaling_RG}. With a
fit for the same region as in the melt case ($N \ge 35$) we obtain a scaling
exponent of $\nu = 0.60$. This is only slightly larger than the theoretical
value $\nu_{\text gs} = 0.588$ for polymers in good solvents due to the finite
chain length. The ratio $R_e^2/R_G^2$ decreases monotonically with chain
length. This indicates that the intramolecular interactions of poly (isoprene)
in the {\it trans} state locally stretch the chains.

To the best of our knowledge only chain size measurements of
\textit{cis}-poly (isoprene) in solution of cyclohexane have been
performed~\cite{davidson87,tsunashima88}. Our data coincides within the order
of magnitude with the extrapolated experimental data of Tsunashima {\it et al.}
to our longest chains. Applying their relation for the gyration radius we
calculate 2.76~nm for a chain length of 100 monomers in $\Theta$-solution.
This is almost exactly our value for the chains of this length in the melt
where also random walk statistics is supposed to apply. Davidson {\it et al.}
measured the radius of gyration of a mixture of {\it cis} and {\it trans} poly
(isoprene) under good solvent conditions by various techniques including wide
angle light scattering for molecular weight range of 112,000~g/mol and
higher.\cite{davidson87} This corresponds to chain lengths of 1650
monomers and longer. Extrapolating the simulation data to the two shortest
chains experimentally available we get 23.7~nm for 112,000~g/mol compared to
17.7~nm experimentally and 26.5~nm for 164,000~g/mol against 19.4~nm
experimentally. Thus, we overestimate the gyration radii by about 35\%.
However, one has to keep in mind that the experimentally investigated system
contains a mixture of the {\it trans} and {\it cis} conformer. The
experimentally observed scaling exponent is 0.545 and therefore lower than our
exponent of 0.6 and also lower than the theoretically predicted exponent of
0.588.

\begin{figure}
  \begin{center}
    \includegraphics[angle=-90,width=8.25cm]{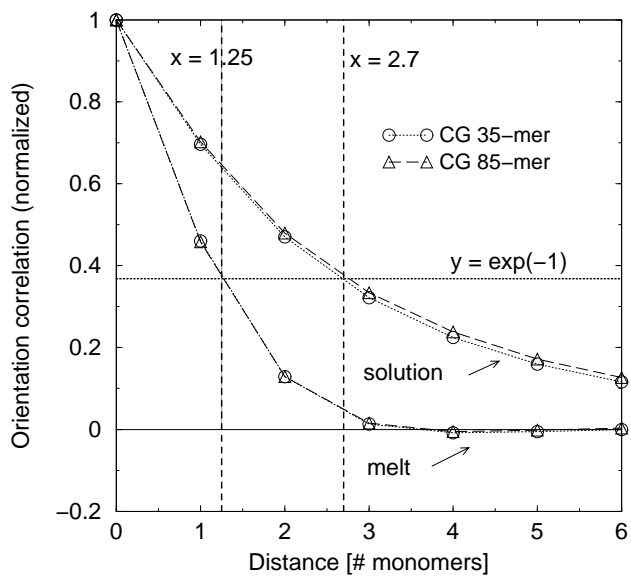}
    \caption{Approximation of the persistence length $l_{\rm p}$ of some
    arbitrarily picked CG poly (isoprene) systems. Using
    eq.~\ref{eq:lpersist}, the decay of the orientation correlation
    function to $y=\exp(-1)$ approximates $l_{\rm p}$. In the melt, this
    yields $l_{\rm p} \approx 1.25l_{\rm b}$ and for the solution,
    $l_{\rm p} \approx 2.7l_{\rm b}$. }
    \label{pip-cg-persist}
  \end{center}
\end{figure}

Next, the persistence length $l_{\rm p}$ of poly (isoprene) was estimated
by the decay of the orientation correlation along the chain backbone. The point
where this correlation arrives at  $y=\exp(-1)$ is a good approximation of
$l_{\rm p}$
\begin{equation}
  \label{eq:lpersist}
  \Big\langle\vec{u}_k(n+j) \; \vec{u}_k(n)\Big\rangle \; = \;
  \exp \left({-\frac{jl_{\rm b}}{l_{\rm p}}}\right) \; .
\end{equation}
Here, $l_{\rm b}$ is the bond length and $\vec{u}_k(n)$ represents the unit
vector along the chain, centered at monomer $n$ and pointing to monomer $k$.
The result is shown in Figure~\ref{pip-cg-persist}. The bonds are
completely decorrelated beyond the forth neighbor for the melt. For
the persistence length, $l_{\rm p} \approx 1.25l_{\rm b}$(= 0.59~nm)
can be estimated. As expected, this is independent of chain lengths,
as all curves fall on top of each other. The same holds for the
solution case. Here the persistence length could be estimated to be
$l_{\rm p} \approx 2.7l_{\rm b}$(=1.27~nm), i.e. more than twice as
large as for the melt. So, the concentration of poly (isoprene) chains
has a strong influence on the intrachain statistics. The dense packing
in the melt favors a quick decorrelation between neighboring chain
monomers.

Finally, we can compare some of the above results to data for a melt of
generic bead-spring chains with bond-angle stiffness~\cite{faller00a}. It has
already been used to successfully map the dynamic behavior of atomistic poly
(isoprene) the meso-scale~\cite{faller02a}, even though it just comprised one
simple bond angle potential in addition to a FENE bead spring interaction for
the intramolecular part:
\begin{equation}
  V_{\rm angle} \; = \; x \cdot
  \left[1+\vec{u}_j\cdot\vec{u}_{j+1}\right] \; k_BT \; .
\end{equation}
The vector $\vec{u}$ corresponds to a normalized bond vector and $x$ is the
force constant to determine the stiffness. The scalar product corresponds to
the cosine of the angle along the backbone. The comparison of static properties
of the two models is presented in Table~\ref{cg-dat-comp}.
One observes that most properties yield similar values, in case of $R_G$,
$R_e^2/R_G^2$, and $l_{\rm p}$ even within statistical error. However, the
end-to-end distances deviate more strongly, showing that the realistic poly
(isoprene) chains are more coiled than generic chains. This is also reflected
in the Kuhn segment length $l_{\rm K}$, which defines the distance beyond
which one expects universal chain behavior:
\begin{equation}
  \label{eq:lkuhn}
   l_{\rm K} \; = \; \frac{\langle\vec{R}^2_e\rangle}
  {R_e^{\text{(max)}}} \; = \; \frac{\langle\vec{R}^2_e\rangle}{(N-1) \; l_b}
   \; = \; C_{\infty} \; l_{\rm b} \; .
\end{equation}
It is significantly shorter for the realistic model compared to the generic
one. A reason for that could be the missing torsional potential in the
latter. Also, the dynamical behavior turns out to differ significantly, as
will be shown in the next section.

\begin{figure}
  \includegraphics[width=8.25cm]{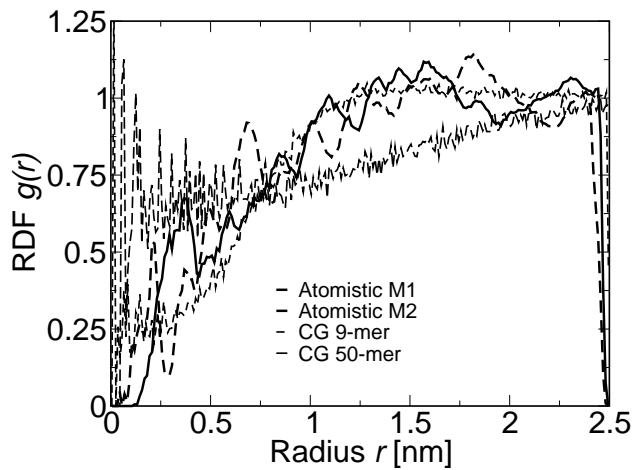}
  \caption{Pair distribution function $g(r)$ for different chain models. We
   show two different atomistic 10-mers (M1 and M2), a mesoscopic 9-mer, and a
   mesoscopic 50-mer.}
  \label{fig:chain-chain}
\end{figure}

Although the mesoscopic model is optimized against the atomistic structure,
some structural differences remain. Comparing the pair distribution
functions of centers-of-mass of whole chains (cf. Figure~\ref{fig:chain-chain})
we see that the mesoscopic model chains show less structure on short distances.
The mesoscopic 9-mers show just a correlation hole whereas the atomistic
10-mers exhibit a structured correlation hole. Moreover we see the strong
difference between a mesoscopic 9-mer and a 50-mer. In the latter case the
correlation hole is much less pronounced. This can be understood as the
bigger a chain becomes the less volume of its ellipsoid of gyration it actually
occupies and the more overlap of such ellipsoids is possible.
\subsection{Dynamical Properties in the melt}
The Rouse model~\cite{rouse53} is widely accepted for the dynamics of short
flexible chains in the melt. The assumptions it is based on are the screening
of excluded volume and that all local intra-chain interactions can be mapped
onto a Kuhn segment length~\cite{kuhn34}. Poly (isoprene) has been
experimentally tested against the Rouse model~\cite{floudas99} and the
agreement was not too good at room temperature. Simulations at the atomistic
model found for the oligomers of length 10 a reasonable description by the
Rouse model at the elevated temperature of $T=413$~K~\cite{faller02a}.

For long chains deviations from the Rouse model are expected due to mutual
topological constraining of chains. Such deviations could then be attributed
to entanglements since recently it has been shown that even a weak
presence of stiffness can dramatically decrease the onset of
entanglement influences~\cite{faller01b}.

\begin{figure}
  \includegraphics[width=8.25cm]{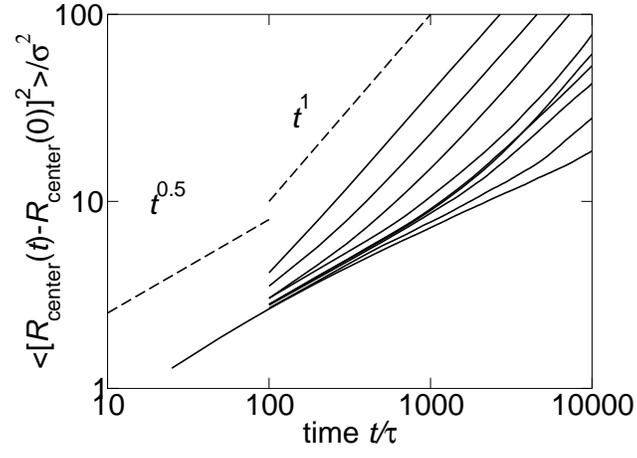}
  \caption{The $g_1$-function, mean-squared displacement of central monomers
    for the melt simulation at chain lengths $N=9,15,20,35,40,45,50,85,120$
    from top to bottom. As guides to the eye we show lines indicating
    $t^{0.5}$ and $t^1$.}
  \label{fig:msdimm}
\end{figure}

Two observables are commonly used to decide on the validity of the Rouse
model: the mean-squared displacements, and the so-called Rouse modes.
Figure~\ref{fig:msdimm} shows the $g_1$ function which is defined as
\begin{equation}
  g_1(t) = \langle (\vec{r}_i(t+t_0) -\vec{r}_i(t_0))^2\rangle_{t_0},
\end{equation}
where $\vec{r}_i(t)$ denotes the position of the central monomer of a chain at
a given time. The central monomer is chosen to suppress end-effects as good as
possible. For short times $(t<\tau_R)$ one expects all curves to fall on top
of each other; $\tau_R$ is the Rouse time with all internal degrees of freedom
are relaxed (below). At such short times the individual monomer does not yet
have to drag the whole chain with it, but an increasing part. At
$t\approx\tau_R(N)$ the different curves leave the master curve as the
increasing neighborhood has reached the chain length. For longer chains
Figure~\ref{fig:msdimm} shows clearly such a Rouse regime where
$g_1(t)\propto t^{0.5}$. Additionally, for all chain lengths $N\le85$ we find
the free diffusion limit $g_1(t)\propto t$ reached and therefore the systems
equilibrated. This shows that the local dynamics is at least Rouse-like
with the connectivity leading to a time dependent diffusion coefficient for
$t<\tau_R$. This in turn leads to the sub-diffusive behavior observed. For
longer chains the local regime lasts longer which clearly hints towards the
connectivity behind the sub-diffusivity.

The actual Rouse model relies on the independence of the Fourier modes of the
chains. We compare the decay of these Rouse modes $\vec{X}_p$
\begin{equation}
  \vec{X}_p=\frac{1}{N}\sum_{i=0}^{N-1}\cos \left(
  \frac{\pi p(i+\frac{1}{2})}{N}\vec{R}_i\right) \; .
\end{equation}
According to theory the curves for the different modes at a given chain length
should collapse if the time and the correlation function are rescaled by the
squared mode index~\cite{doi86}. For intermediate chain lengths we find this
scaling to work reasonably well. For longer chain lengths, however, the higher
modes are apparently faster than predicted by the Rouse model. This is first
hint towards entanglements. It has been observed in simulations of other
polymer models that especially the lower modes are likely to be the first to
be influenced by entanglements~\cite{puetz99,faller02a}.

\begin{figure}
  \includegraphics[angle=-90,width=8cm]{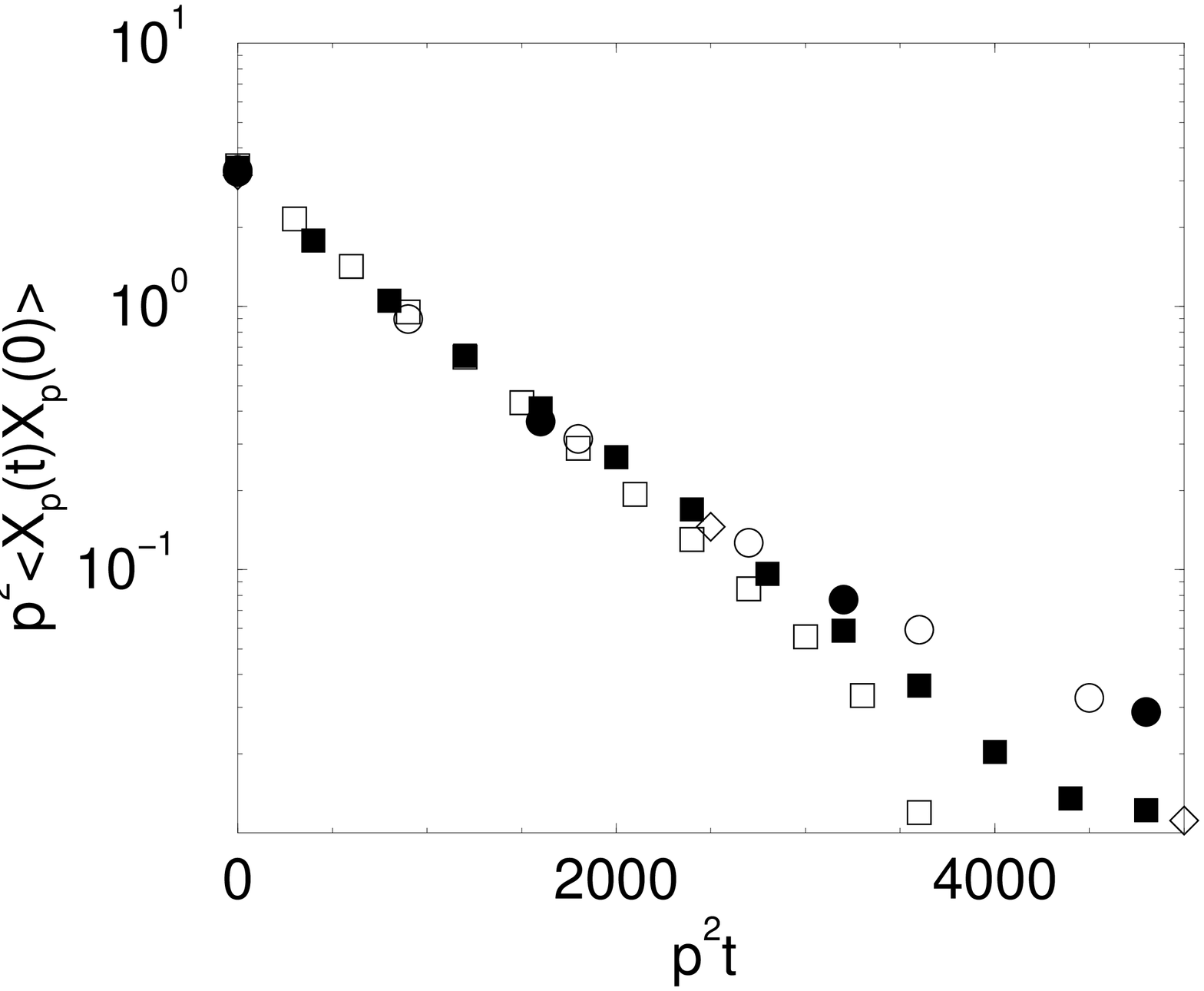}
  \includegraphics[angle=-90,width=8cm]{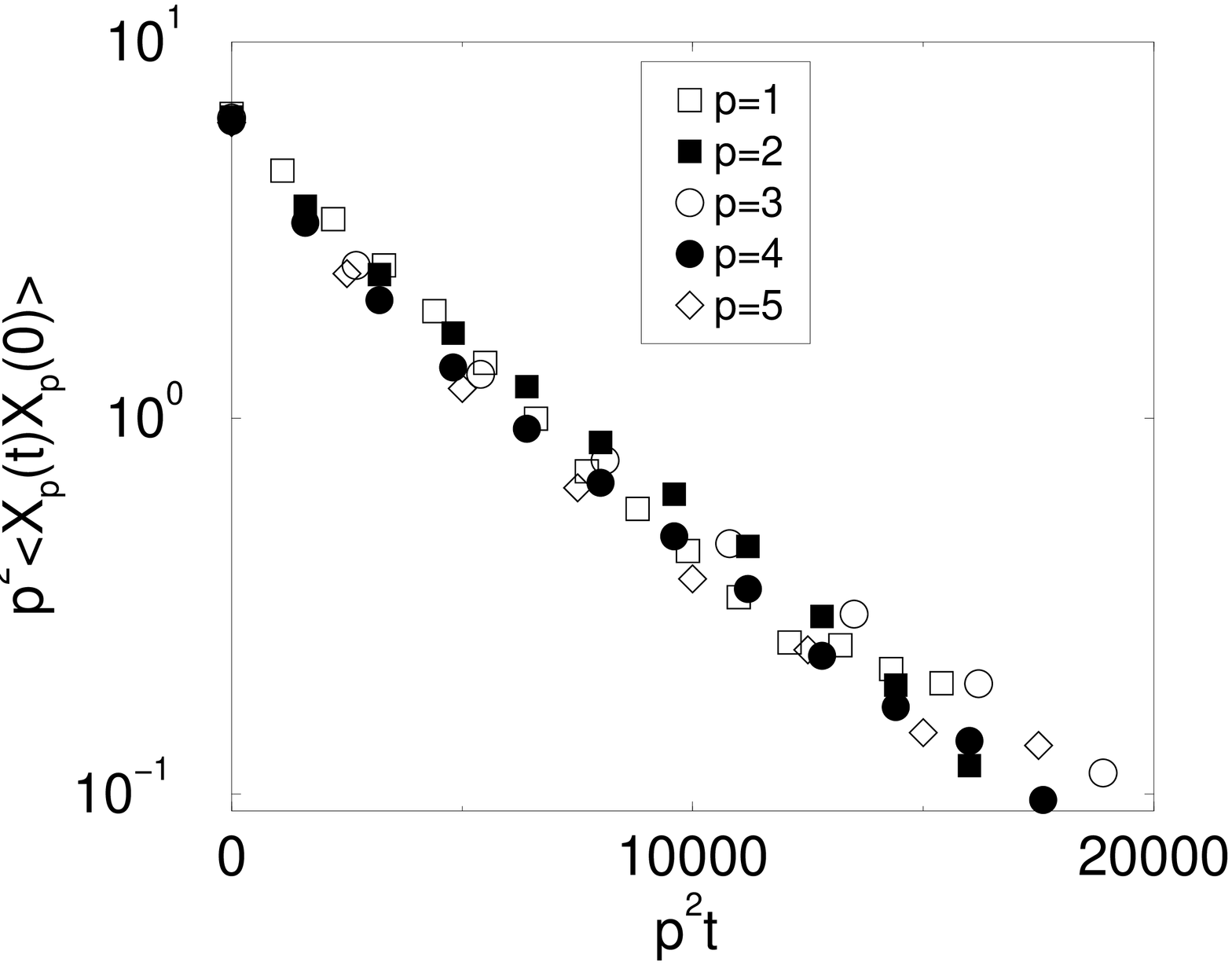}
  \includegraphics[angle=-90,width=8cm]{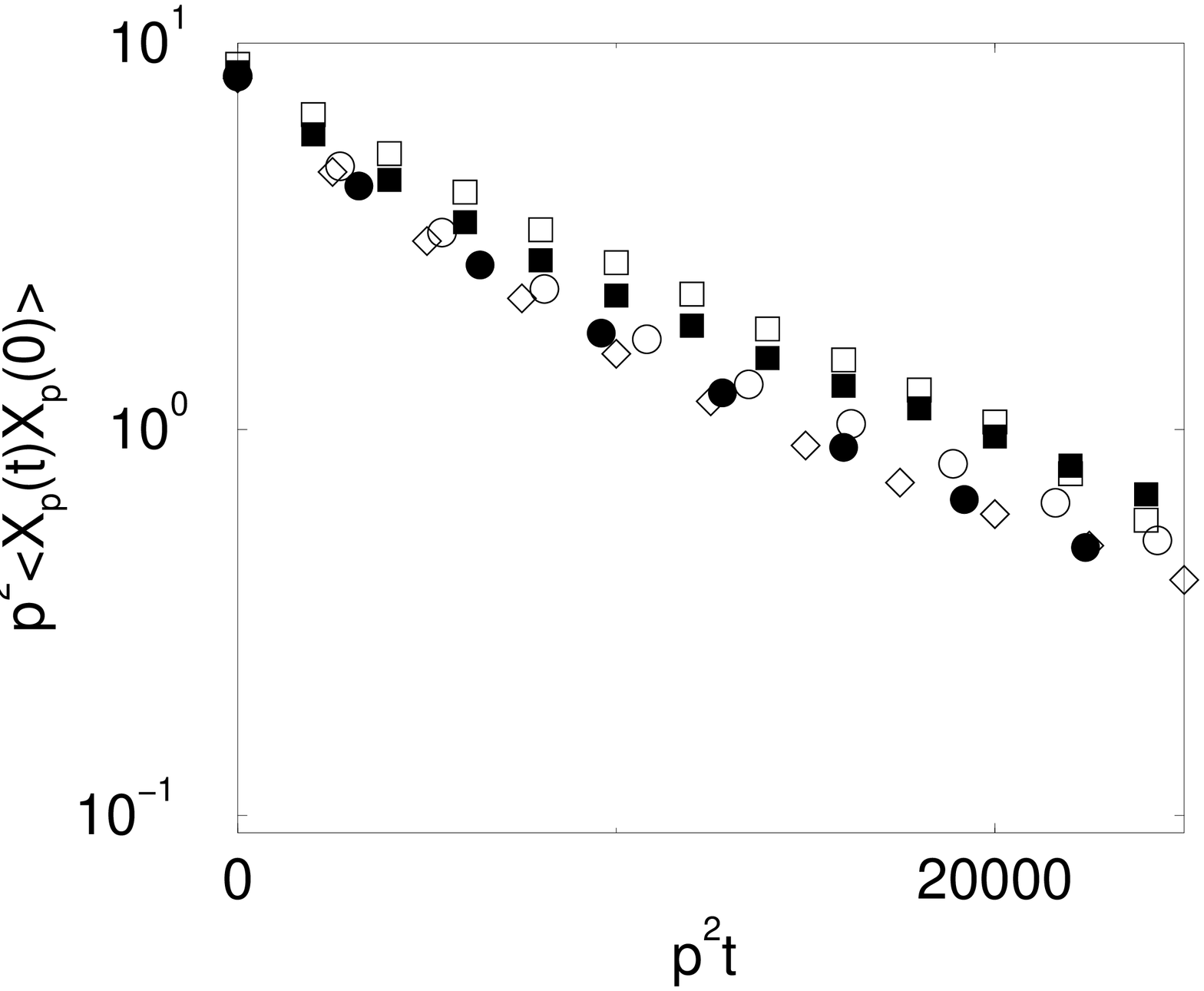}
  \includegraphics[angle=-90,width=8cm]{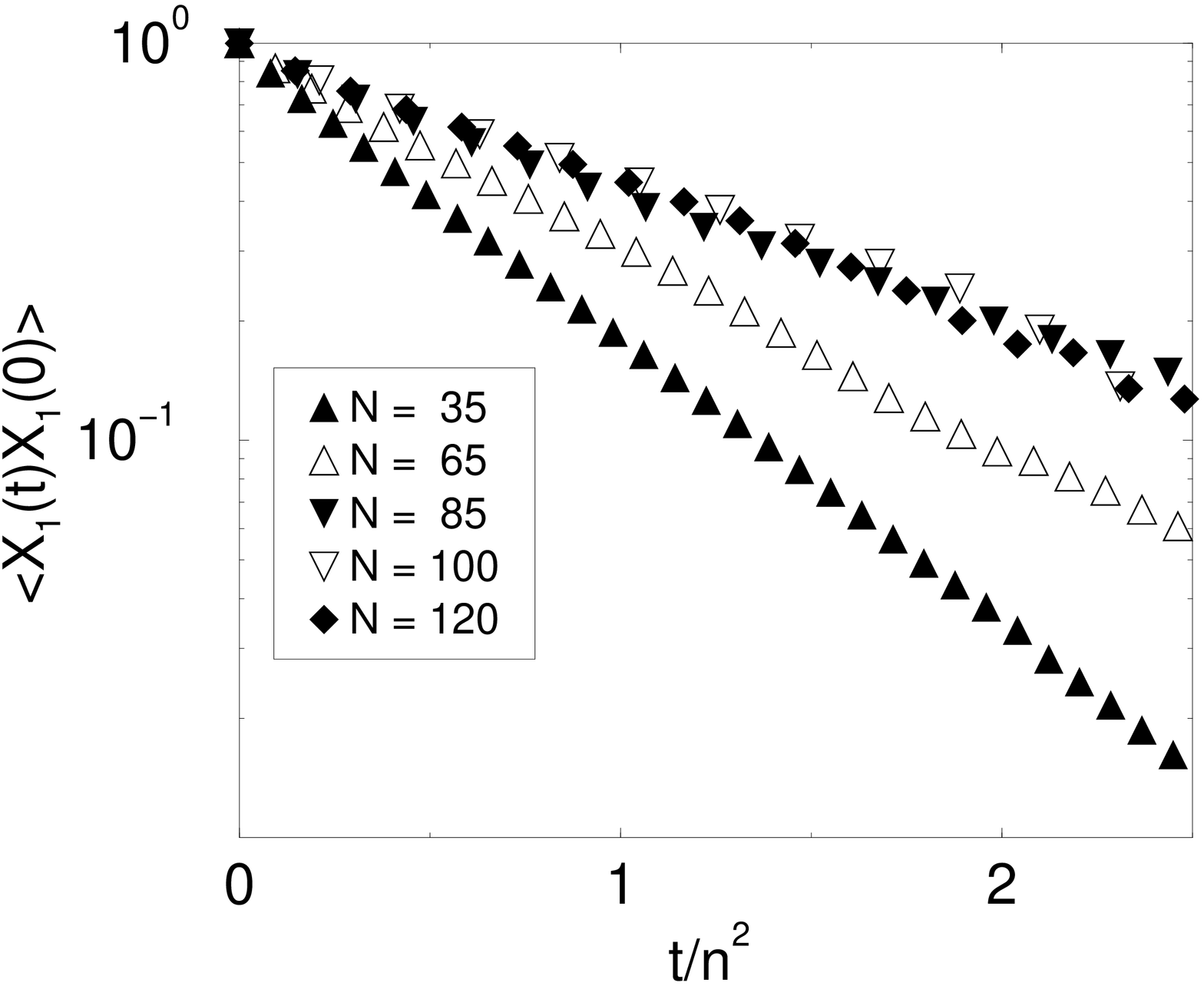}
  \caption{Decay of the autocorrelation functions of the Rouse Modes for
  different chainlengths in Rouse scaling. Figure a) $N=35$ b) $N=65$,
  c) $N=85$. The symbols are described in figure b. d) Comparison of the decay
  of the autocorrelation function for different chain lengths.
 }
  \label{fig:Rousemodes}
\end{figure}

Table~\ref{tab:RT} shows the Rouse time $\tau_R$ derived by an exponential
fit to the first Rouse-Mode $\vec{X}_1$. We clearly see finite chain length
effects. For longer chains $N\ge 100$ these effects start to become weaker.
Figure~\ref{fig:Rousemodes} compares the scaling for several chain lengths.
This suggests that poly (isoprene) melts at room temperature do not
completely behave according to Rouse theory.

We are not yet able to reach into the entanglement regime. Especially for the
longest chains the scaling of the Rouse modes for different chain lengths onto
each other apparently holds. It is interesting to note that we see finite chain
length effects for chains of moderate length. Figure~\ref{fig:Rousemodes}d is
a way to tell oligomers from polymers as the self-similarity is obeyed in the
polymer case. For oligomers the Rouse friction is still chain length dependent.
This was not found in a recent study of model chains only incorporating
stiffness. In that case the Rouse model was obeyed almost from the very
beginning ($N>15$). This is another important finding in the view of modeling
long oligomers or short polymers. For chains with chemical identity we find a
regime where the chain nature is already important but the generic polymer
models are not yet valid.
\section{Conclusions}
Atomistic simulations of trans-poly (isoprene) in the melt and in solution
have been simulated and independently coarse-grained to models with one
interaction center per monomer. The local scale models are verified against
experimental data where available. In the melt a slowdown of the
dynamics is found as expected. For the solution we were able to equilibrate
oligomers of length 15 monomers until the end-to-end reorientation correlation
function decayed to less than about 30\% of its initial value. Note, that we
did not need heavy hardware resources to generate the statistical precision we
present here.

The two independently coarse-grained models for solution and melt respectively
are distinctly different. It is impossible to represent a chain in such
strongly different environments on the coarse grained level with a generic
potential. This is especially true as we aimed to get rid of the solvent
completely in the solution case and melt and solution are very well known to
behave strongly different in the long chain limit.

The coarse-grained models could be compared against well-known long polymer
scaling theory. Additionally, comparison to experiments yielded a difference
of 35\%. The Rouse behavior is for this model recovered for chains of more
than around 80 monomers. In summary, we find that the proposed coarse graining
procedure allows us to produce models which are reliably reproducing
properties of real polymers. We were not yet able to reach into the reptation
regime.
\section*{Acknowledgements}
Many fruitful discussions with Florian M\"uller-Plathe and Hendrik Meyer are
gratefully acknowledged.  RF wants to thank the Emmy-Noether program of the
DFG (German Research Foundation) and DR wants to thank the BMBF (German
Department of Research) Competence Center in Materials Simulation for financial
support.

\bibliography{standard,cg6}
\bibliographystyle{macromolecules}
\clearpage

\clearpage

\clearpage
\begin{table}
    \caption{Angles (equilibrium values $\phi_{0}$ and potential
    strength $k$) and bond lengths $l_{b}$ for atomistic melt
    simulations of poly (isoprene). The bond lengths are constrained
    using SHAKE~\cite{ryckaert77}.}  \[
    \begin{array}{crc|cD{.}{.}{-1}} \hline \text{angle} &
    \mc{\phi_{0}} & k \text{[kJ/(mol*rad}^2)] & \text{bond} & l_{b}
    \text{[nm]} \\ \hline \opC_1-\opC_2-\opC_3 & 128.7 & 250 &
    \opC_1-\opC_2 & 0.150 \\ \opC_2-\opC_3-\opC_4 & 124.4 & 250 &
    \opC_2=\opC_3 & 0.1338 \\ \opC_2-\opC_3-\opC_5 & 120.2 & 250 &
    \opC_3-\opC_4 & 0.151 \\ \opC_4-\opC_3-\opC_5 & 115.4 & 250 &
    \opC_3-\opC_5 & 0.1515 \\ \opC_3-\opC_5-\opC_1 & 114.5 & 250 &
    \opC_5-\opC_1 & 0.155 \\ \opC_5-\opC_1-\opC_2 & 112.7 & 250 &
    \opC-\opH & 0.109 \\ \opC-\opC_{\text{sp3}}-\opH & 109.5 & 250 & &
    \\ \opC_1-\opC_2-\opH & 114.4 & 250 & & \\ \opC_3-\opC_2-\opH &
    114.4 & 250 & & \\ \opH-\opC-\opH & 109.5 & 250 & & \\ \hline
    \end{array} \] \label{tab:ffangbnd}
\end{table}
\begin{table}
    \caption{Force-field parameters of the non-bonded interactions
    for atomistic melt simulations of poly (isoprene). $m$ is the atom
    mass, $\sigma$ the interaction radius, and $\epsilon$ the
    interaction strength. }
    \[
      \begin{array}{cD{.}{.}{-1}D{.}{.}{-1}D{.}{.}{-1}} \hline
      \text{atom} & \mul{m\text{[amu]}} & \mul{\sigma\text{[nm]}} &
      \mul{\epsilon\text{[kJ/mol]}}\\ \hline
      \opC_{\text{sp2}} & 12.01 & 0.321 & 0.313\\
      \opC_{\text{sp3}} & 12.01 & 0.311 & 0.313\\
      \opH & 1.00782 & 0.24 & 0.2189\\ \hline
    \end{array}
    \]
    \label{tab:ff}
\end{table}
\clearpage
\begin{table}
  \begin{center}
    \caption{Force-field parameters of the dihedral angles for atomistic melt
    simulations of poly (isoprene). The identifying numbers are found in
    figure~\ref{polyisoprene}}
    \begin{tabular}{crcrcc}
      \hline
      torsion & \mult{equilibrium angle} & \mult{barrier} & periodicity \\
      & \mult{$\tau_0$ [degrees]} & \mult{$k_{\tau}$[kJ/mol]} & $i$ \\
      \hline
      1 &   0 & &  5.2 & & 1 \\
      1 &   0 & & -7.4 & & 2 \\
      1 &   0 & & 10.0 & & 3 \\
      2 & 180 & &  9.7 & & 1 \\
      2 & 180 & & 14.1 & & 3 \\
      3 &   0 & &-21.1 & & 1 \\
      3 &   0 & &-12.3 & & 2 \\
      3 &   0 & &  0.5 & & 3 \\
      \hline
    \end{tabular}
    \label{tab:torspot}
  \end{center}
\end{table}
\begin{table}
    \caption{Summary of the characteristics of the atomistic melt simulations.}
  \[
    \begin{array}{ccrrD{.}{.}{-1}}
      \hline
      \text{system} & T\text{[K]} & t_{\text{sim}}\text{[ps]}
      & \frac{M_w}{M_n} & \mc{\rho\text{[kg/m}^3]} \\
      \hline
      M1 & 300 & 1184 & 1.00 & 890\\
      M2 & 300 & 2012 & 1.05 & 917.4\\
      M3 & 300 & 1737 & 1.05 & 916.8\\
      M3 & 413 &  792 & 1.05 & 826\\
      \hline
    \end{array}
    \]
    \label{tab:simlen}
\end{table}
\begin{table}
  \begin{center}
    \caption{Thermodynamic and static properties of the solution. Details of
      the different systems and the numbers with which they are referenced in
      the following. $N_P$ is the number of oligomers (C$_{75}$H$_{122}$) and
      $N_C$ the number of cyclohexane molecules. $c$ is the concentration in
      weight \% polymer. $t_{\text{sim}}$ is the simulated time for the
      systems.  }
    \begin{tabular}{ccccccccc}
      \hline
      system & $N_P$ & $N_C$ & c & \mc{$\rho$}\text{[kg/m}$^3$] &
      $t_{\text{sim}}$ [ns] & $R_H$ [nm] & $R_G$ [nm] & $R_e^2/R_G^2$ \\
      \hline
      S0 & 0 & 500 & 0.0\% & 756.3 & 1.0  & -- & -- & --\\
      S1 & 1 & 250 & 4.6\% & 764.2 & 11.25 & 1.33 $\pm$ 0.12 & 1.21 $\pm$ 0.20
      & 6.1 \\
      S2 & 1 & 500 & 2.4\% & 757.5 &  5.58 & 1.40 $\pm$ 0.05 & 1.26 $\pm$ 0.10
      & 6.0 \\
      S3 & 2 & 500 & 4.6\% & 762.5 &  7.81 & 1.34 $\pm$ 0.10 & 1.23 $\pm$ 0.20
      & 6.5 \\
      S4 & 2 & 250 & 8.9\% & 768.2 & 11.56 & 1.37 $\pm$ 0.09 & 1.29 $\pm$ 0.15
      & 7.2 \\
      \hline
    \end{tabular}
    \label{tab:systems}
  \end{center}
\end{table}
\begin{table}[hp]
  \begin{center}
    \caption{Data of the mesoscopic Brownian Dynamics poly (isoprene) melt
      simulations. The pressure is stated in reduced units~\cite{allen87}. All
      systems are simulated at a temperature of $T=300$~K and at a density of
      $\rho_p=7.25$ monomers/nm$^3$ and correspond to the atomistic simulation
      M1. The static properties come with a standard error of $\approx 5\%$,
      the dynamic properties with an error of $\approx 15\%$.}
     \begin{tabular}{l||r|r|r|r|r|r|r|r|r|r|r}
chain length $N$       &  9  & 15  & 20  & 35 & 40 & 45 & 50 & 65 & 85 & 100 &
120\\ \hline
number of chains $N_c$ & 100 & 150 & 150 & 100 & 150 & 150 & 80 & 130 & 50 &
120 & 100\\
pressure $p^*$         & $\pm$0.0 & -0.01 & -0.04 & -0.15 & -0.07 & -0.08 &
-0.08 & -0.08 & -0.09 & -0.09 & -0.09\\
simulation time      & 3.0 & 4.0 & 4.0 & 7.2 & 6.0 & 6.0 & 8.0 & 6.0 & 8.0 &
6.0 & 6.0\\
~[$10^6$ time steps]   & & & & & & \\
$R_G$ [nm]         & 0.75 & 1.03 & 1.20 & 1.63 & 1.75 & 1.87 & 1.97 & 2.25 &
2.63 & 2.85 & 3.14\\
$R_H$ [nm]         & 1.11 & 1.19 & 1.27 & 1.50 & 1.60 & 1.65 & 1.69 & 1.98 &
1.98 & 2.21 & 2.42\\
$R_G/R_H$                & 0.68 & 0.86 & 0.94 & 1.09 & 1.10 & 1.13 & 1.17 &
1.14 & 1.33 & 1.29 & 1.30\\
$R_e^2/R_G^2$       & 6.0 & 6.0 & 6.0 & 6.0 & 6.0 & 6.0 & 6.0 & 6.0 & 6.0 &
6.0 & 6.0\\
$D_{\rm centre}$ [10$^{-6}$cm$^2$/s] & 16.6 & 10.7 & 6.7 & 4.2 & 2.5 & 2.3 &
2.9 & 1.3 & 1.3 & 0.8 & 0.6\\
$D_{\rm com}$ [10$^{-6}$cm$^2$/s]    & 16.0 & 10.9 & 6.7 & 4.1 & 2.5 & 2.3 &
2.8 & 1.3 & 1.2 & 0.8 & 0.6\\
    \end{tabular}
   \label{tab:pip_melt}
  \end{center}
\end{table}
\begin{table}[htbp]
  \begin{center}
    \caption{Static properties (with standard errors of $\approx 5\%$) as
      obtained from Monte-Carlo simulations of mesoscopic poly (isoprene)
      solutions. The system density and temperature corresponds to the parent
      atomistic systems S1 and S3.}
    \begin{tabular}{c||c|c|c|c|c|c|c}
       chain length $N$   &  9  & 15  & 35 & 50 & 85 & 100 & 120\\ \hline
       $R_G$ [nm]         & 0.91 & 1.33 & 2.37 & 2.92 & 4.11 & 4.48 & 4.98 \\
       $R_H$ [nm]         & 1.23 & 1.40 & 1.98 & 2.32 & 3.06 & 3.31 & 3.63 \\
       $R_G/R_H$          & 0.74 & 0.95 & 1.20 & 1.26 & 1.34 & 1.35 & 1.37 \\
       $R_e^2/R_G^2$      & 7.3 & 7.1 & 6.7 & 6.5 & 6.5 & 6.4 & 6.4 \\
    \end{tabular}
    \label{tab:pip_solution}
  \end{center}
\end{table}
\begin{table}[hbtp]
  \begin{center}
    \caption{Results of some static properties for a realistic meso-scale
      model of poly (isoprene) in a melt and a more generic bead-spring model
with
      bond-angle stiffness. $R_G$ is the radius of gyration, $R_e$ the
      end-to-end distance of the chain, $l_{\rm p}$ corresponds to the
      persistence length and $l_{\rm K}$ to the Kuhn segment length.}
    \begin{tabular}{c||c|c|c|c|c}
Melt Model & $R_G$ & $R_e^2$ & $R_e^2/R_G^2$ & $l_{\rm p}$ & $l_{\rm K}$\\
& [nm] & [nm$^2$] & & [monomers] & [nm]\\\hline\hline
$x=1.5$, $N=20$~\cite{faller00a} & 1.28 & 10.3 & 6.3 & 1.2 & 1.2\\
CG PI, $N=20$ & 1.20 & 8.7 & 6.0 & 1.25 & 1.0\\\hline
$x=1.5$, $N=50$~\cite{faller00a} & 2.20 & 29.5 & 6.2 & 1.3 & 1.3\\
CG PI, $N=50$ & 1.97 & 23.3 & 6.0 & 1.25 & 1.0\\
    \end{tabular}
    \label{cg-dat-comp}
  \end{center}
\end{table}
\begin{table}
  \caption{Fits of the Rouse time using the decay of the first Rouse mode for
    different chain lengths.}
  \begin{tabular}{rrr}
   $N$ & $\tau_R$ & $\tau_R/N^2$\\
   35 &   743 & 0.60 \\
   50 &  1946 & 0.78 \\
   65 &  3722 & 0.88 \\
   85 &  9300 & 1.29 \\
  100 & 13800 & 1.38 \\
  120 & 17700 & 1.23 \\
  \end{tabular}
  \label{tab:RT}
\end{table}
\end{document}